\documentclass[two column,showpacs]{revtex4}
\pdfoutput=1
\usepackage[T1]{fontenc}
\usepackage{graphicx}
\usepackage{amsmath}
\usepackage{amssymb}
\usepackage{amsfonts}
\usepackage{color}
\usepackage{appendix}
\usepackage{caption}
\usepackage{subcaption}

\def\a{\alpha}
\def\b{\beta}
\def\g{\gamma}
\def\e{\varepsilon}
\def\d{\delta}

\def\k{\kappa}

\def\m{\mu}

\def\t{\tau}
\def\n{\nu}

\def\S{\Sigma}

\def\D{\Delta}
\def\o{\omega}

\def\be{\begin{equation}}
\def\ee{\end{equation}}
\def\bea{\begin{eqnarray}}
\def\eea{\end{eqnarray}}
\def\llang{\langle\langle}
\def\rrang{\rangle\rangle}
\def\nn{\nonumber}
\def\lb{\label}

\begin{document}
	
	\title{Impurity effects in twisted carbon nanotubes}
	
	\author{Yuriy G. Pogorelov}%
	\email{ypogorel@yahoo.com}
	\affiliation{IFIMUP,~Departamento~de~F\'{i}sica e Astronomia,~Universidade~do~Porto,~Porto,~Portugal,}

        \author{Volodymyr Turkowski}%
	\email{Volodymyr.Turkowski@ucf.edu}
	  \affiliation{Department~of~Physics,~University~of~Central~Florida,~FL~32816,~USA} 

      \author{Vadim M. Loktev}%
	\email{vloktev@bitp.kyiv.ua}
	\affiliation{N.N.~Bogolyubov~Institute~for~Theoretical~Physics,~NAS~of~Ukraine,~Kyiv,~Ukraine,}

\begin{abstract}
We consider electronic spectra of twisted carbon nanotubes and their perturbation by impurity atoms absorbed at different positions on 
nanotube surface within the framework of Anderson hybrid model. A special attention is given to the cases when 1D Weyl (massless Dirac) 
modes are present in the nanotube spectrum and their hybridization with localized impurity states produces, with increasing impurity 
concentration $c$, the onset of a mobility gap near the impurity level and then, at further growth of $c$, opening of a narrow range of 
delocalized states within this mobility gap. Such behaviors are compared with similar effects in the previously studied nontwisted carbon nanotubes. Some possible practical applications are discussed.
\end{abstract}
	
\date{\today}
\keywords{carbon nanotubes, impurity adatoms, spectrum restructuring}
\maketitle

\section{Introduction}\lb{intr}
	
Carbon nanotubes (CNTs) \cite{Iijima2002, Nevidomskyy, Charlier2007, Wakabayashi2010, Devolder2013, Inagaki2014} are a broad class
of 1D-like structures formed on the basis of 2D graphene \cite{Geim2005, Geim2009, Tomanek_Book}, with an important possibility for
their electronic spectra to include 1D Weyl modes (WMs) \cite{Charlier2007}. This importance is due to the combination of linear 
low-energy dispersion of charge carriers with their constant 1D density of states, giving promises for very sensitive controls over 
conductive, optical and other physical properties of these systems. 

In analogy with known conductivity tuning in semiconductors, such controls can be realized by introduction of impurities (dopants) but 
specifics of CNTs is in variety of their dimensional and topological parameters that can qualitatively change the phase structure of 
low energy eigenstates and their response to external factors. These issues are broadly discussed for 1D carbon nanoribbons (CNRs) and 
for two basic CNT classes: zigzag nanotubes (ZNTs) and armchair nanotubes (ANTs). But there is a much wider general field of CNT 
structures between these two limits, known as twisted nanotubes (TNTs) \cite{Saito, Iijima2002, Nevidomskyy}, especially rich in their 
topological parameters and rising interest to their conducting behaviors under impurity doping. The present study is focused on these
behaviors in function of TNT geometries and in comparison with the ZNT and ANT references.

Various technological methods for TNT synthesis are now in active development \cite{Munoz, Lim, Rathinavel2021, Lou}, also their 
electronic structure is being studied theoretically, including the first principles approaches \cite{Kato}.

The main purpose of the present work is the theoretical description of electronic dynamics in various TNTs, differing by their dimension 
(diameter) and topology (twisting angle), and in establishing the conditions for their maximum sensitivity to dopants and to external 
controls.

The following presentation begins from definition of lattice structure parameters in TNTs (Sec. \ref{Geo}), 
then continued by the 
analysis of their electronic spectra in the tight-binding approximation with special emphasis on the WMs parameters (Sec. \ref{WM}). 
The most important study of impurity effects on WMs is done in the T-matrix approximation (Sec. \ref{Imp}) and then completed by its 
validity justification (Sec. \ref{Bey}). The final discussion of the obtained results and their possible practical applications is given 
in Sec. \ref{Disc}. Some supplementary materials are presented in Appendices \ref{Mat}, \ref{Sdet}, \ref{Fv}.

\section{Geometry of twisted nanotubes}\lb{Geo}
Each carbon nanotube (CNT) can be obtained by rolling of a carbon nanoribbon (CNR) cut out from the graphene plane so that its edges 
seamlessly match at rolling. Usually, CNRs are classified by their edges, the simplest types being a zigzag nanoribbon (ZNR), with its 
edges along the graphene basic vector $\bold{a}_1$, and an armchair nanoribbon (ANR), with those along the graphene unit cell diagonal 
$\bold{a}_1 + \bold{a}_2$ (with the lattice constant $|\bold{a}_1| = |\bold{a}_2| = a \approx 2.46 \mathring{A}$). But CNTs (having no 
edges) are classified by the normals to their axes, thus an ANR gets rolled into a zigzag nanotube (ZNT) and a ZNR does into an armchair 
nanotube (ANT). 

More generally, any CNR (and the unfolded CNT) is characterized by its rectangular unit cell based on the chiral (transversal) vector:
\be
{\bold C}_h = n\bold{a}_1 + m\bold{a}_2,
\lb{eq1}
\ee
with mutually prime chiral indices $n$ and $m$ and on the translational (longitudinal) vector:
\be
\bold T = \frac{2n + m}d\bold{a}_2 - \frac{2m + n}d \bold{a}_1,
\lb{eq2}
\ee
where $d$ is the greatest common divisor of $2n + m$ and $2m + n$. Without loss of generality, we take $n \geq m$ in what follows. The total 
number of sites in the unit cell:
\be
M = \frac{4C_h T}{\sqrt 3 a^2} = \frac{4}{d}(n^2 + mn + m^2),
\lb{eq3}
\ee
is always even (by equivalence of graphene two sublattices) and the full CNT length $NT$ is taken much longer of the longitudinal period $T$. 

One limiting CNT type is ZNT ($m = 0$, $d = n$) with the transversal period $C_h = a$ (along the circumference of $na$) and the longitudinal period 
$T = \sqrt 3 a$. Another limit is ANT ($m = n$, $d = 3n$) with the transversal period $C_h = \sqrt 3a$ (along the circumference of $n\sqrt 3 a$) 
and the longitudinal period $T = a$. 
 \begin{figure}
\centering
  \includegraphics[width=8cm]{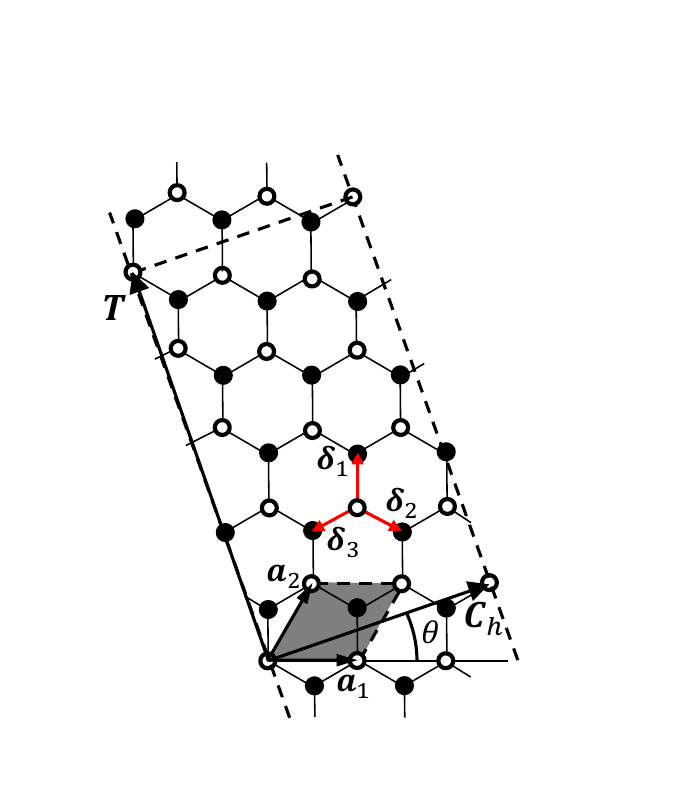}
  \caption{The (2,1) TNT unfolded unit cell based on vectors ${\bold C}_h = 2{\bold a}_1 + {\bold a}_2$ and ${\bold T} = -4{\bold a}_1 
  + 5{\bold a}_2$ and the graphene unit cell (shadowed) based on vectors ${\bold a}_1$ and ${\bold a}_2$ with two atomic sites (white and 
  black circles). The nearest-neighbor vectors $\boldsymbol{\d}_{1,2,3}$ (red) and the chiral angle $\theta$ are indicated.}
  \label{fig1}
\end{figure}

\begin{figure}
  \centering
  \includegraphics[width=10cm]{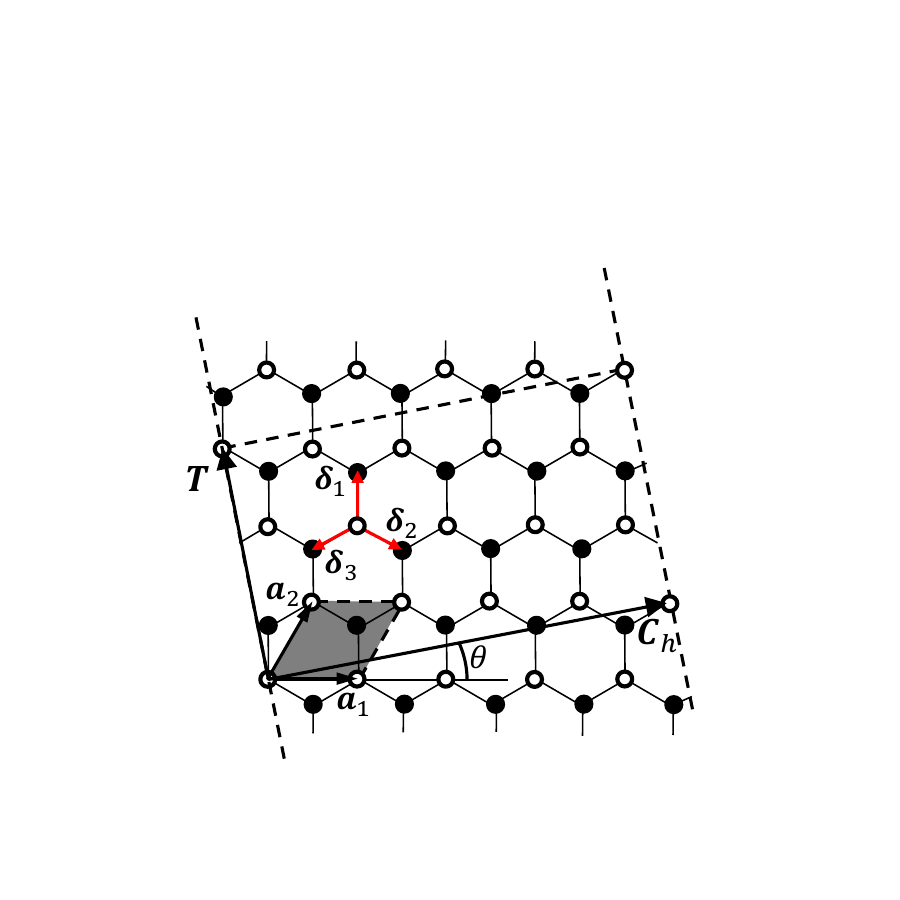}
  \caption{The (4,1) TNT unfolded unit cell based on vectors ${\bold C}_h = 4{\bold a}_1 + {\bold a}_2$ and ${\bold T} = -2{\bold a}_1 
  + 3{\bold a}_2$  differs from that of the (2,1) one by its smaller chiral angle $\theta$ while the total number of atoms in both cases 
  is the same: $M = 28$.}
  \label{fig2}
\end{figure}

All other $n,m$ indices refer to TNTs with their periods $C_h$ and $T$ by Eqs. \ref{eq1}, \ref{eq2}. Each TNT can also be defined by its diameter 
$D = \n C_h/\pi$, with a natural $\n$ number of transversal periods along its circumference, and the chiral angle:
\be
\theta = \arctan \frac{\sqrt 3 m}{2n + m},
\lb{theta}
\ee
between the vectors ${\bold C_h}$ and ${\bold a}_1$, with its values $0 < \theta < \pi/6$ (between the ZNT and ANT limits). Some examples 
of TNT are presented by their unfolded unit cells in Figs. \ref{fig1}, \ref{fig2}. Each unit cell can be labeled by its longitudinal number 
$p$ and its chiral number $j$ and each atomic site in this cell can be labeled by its $s$ index.

\section{Electronic types of twisted nanotubes}
\lb{WM} 
Physically, two basic CNT electronic types can be denoted:

$\ast$ insulating, presenting a finite energy gap around the Fermi level, and

$\ast$ metallic, with their spectrum including pairs of 1D WMs linearly crossed at zero-energy Fermi level. 

Our next focus is on how CNT physical properties are related to their $n,m$ indices. Thus, it is known that any metallic CNT structure 
should satisfy the basic condition \cite{Charlier2007}:
\be
n - m = 0\, ({\rm mod}\,3),
\lb{met}
\ee
that is, $n = m \,+\,3l$ with a whole $l$. Its simplest realizations are readily found in all ANTs ($n = m$) and in the (3$l$, 0) ZNTs. The condition 
by Eq. \ref{met} for mutually prime $n$ and $m$ readily eliminates all $m = 0 \,({\rm mod}\,3)$ values and also leads to the single non-trivial 
value of $d = 3$. 

The latter can be proved from the contrary: if another common divisor $d'$ of $2n + m$ and $2m + n$ (besides $d = 3$) is supposed to 
exist, it should be the greatest common divisor of $(2n + m)/3 = m + 2l$ and $(2m + n)/3 = m + l$, that is: $d' = (m + 2l)/(m + l)$. But this 
ratio also reads as $d' = 1 + l/(m + l)$ which evidently satisfies $1 < d' < 2$ for all $m$ and $l$, so cannot be a non-trivial whole number. 

Then the longitudinal period by Eq. \ref{eq2} for all metallic TNTs is presented as:
\be 
T = \sqrt{\frac{n^2 + m^2 + mn}{3}} = \frac{C_h}{\sqrt 3},
\lb{CT}
\ee
and the constant ratio of these two periods, $T/C_h = 1/\sqrt 3$ is important in the following analysis of TNT electronic dynamics. 

Otherwise, for all insulating TNTs, that is for:
\be
n - m = \pm 1\,(\rm mod\, 3),
\lb{ins}
\ee
it can be proved that $2n + m$ and $2m + n$ turn here mutually prime (only permitting the trivial $d = 1$ value). Alike the above for  
metallic ones, suppose the contrary of a non-trivial $d = (2n + m)/(2m + n)$ to exist, then it is rewritten here as: 
\[d = \frac{3m + 6l \pm 2}{3m + 3l \pm 1} = 1 + \frac{3l \pm 1}{3m + 3l \pm 1},\]
which is also restricted to $1 < d < 2$, thus not being a whole number. Then, readily, the ratio of TNT periods stays constant 
but inverted {\it vs} the metallic case: $T/C_h = \sqrt 3$. 

These relations of periods for the two TNT types are illustrated by the examples in Figs. \ref{fig1}, \ref{fig2}. 

\section{Quasiparticle states in TNTs}

Generally, the local Fermi operators $a_{p,j,s}$ for all the $s$ sites ($1 \leq s \leq M$) in the $p,j$ unit cell can be combined into the local 
$M$-spinor:
\be
a_{p,j} = \left(\begin{array}{c}
		a_{p,j,1}  \\
		a_{p,j,2}  \\
		\dots  \\
		a_{p,j,M}  \\
	\end{array}\right).
\lb{asp}
\ee
Then the TNT translational invariance with the $C_h, T$ periods by Eqs. \ref{eq1}, \ref{eq2} suggests, in similarity to the previously analyzed 
ZNT and ANT cases, the Fourier expansion of this spinor components in quasi-continuous translational momenta $-\pi < k < \pi$ and in discrete 
transversal wave numbers $q$:
\bea
&& a_{p,j,s} = \frac 1{\sqrt{2\pi M\n}}\sum_{q = 0}^{\n - 1}\int_{-\pi}^\pi dk \exp \left[i\left(k t_{p,j,s}\right.\right.\nn\\
&& \qquad\qquad\qquad\qquad\qquad + \left.\left.\frac{2\pi q}{\n} c_{j,s}\right)\right]\a_{k,q,s},
\lb{apj}
\eea
where $t_{p,j,s}$ and $c_{j,s}$ are respectively longitudinal (in $T$ units) and transversal (in $C_h$ units) coordinates of the $p,j,s$-site. Alike 
that by Eq. \ref{asp} for local operators, the $\a_{k,q,s}$ wave operators can also be combined into the wave $M$-spinor $\a_{k,q}$. 

The next treatment of TNT electronic dynamics is done in the approximation of nearest-neighbor hopping between carbon $\pi$-orbitals with the same 
hopping parameter $\g \approx 2.8$ eV as in 2D graphene \cite{Geim2009}, equivalent to the zone-folding approximation \cite{Samsonidze} (see its 
plausibility below in Sect. \ref{Disc}). In this approximation, the TNT Hamiltonian is written in terms of wave spinors as:
\be
H = \g\sum_{q = 0}^{\n - 1}\int_{-\pi}^\pi dk \,\a_{k,q}^\dagger\hat H(k,q)\a_{k,q}.
\lb{HZ}
\ee

Here the $M$$\times$$M$ Hermitian matrix $\hat H(k,q)$ has only three non-zero elements in each of its rows and columns, of the form $z_\m(k,q) = 
{\rm e}^{i\varphi_\m(k,q)}$, $\m = 1,2,3$ (or their complex conjugates $z_\m^\ast(k,q)$). Their phases: 
\bea
&& \varphi_1(k,q) = \frac{a}{\sqrt 3}\left(\frac k T \cos\theta + \frac{2\pi q}{C_h\n}\sin\theta\right),\nn\\
&& \varphi_2(k,q) = -\frac{a}{\sqrt 3}\left[\frac k T\sin(\frac \pi 6 + \theta) - \frac{2\pi q}{C_h\n}\cos(\frac \pi 6 + \theta)\right],\nn\\
&&\qquad\qquad \varphi_3(k,q) = -\varphi_1(k,q) - \varphi_2(k,q),
\lb{uvw}
\eea
result from the $k,q$-Fourier transformed products of local operators at nearest-neighbor sites separated by the corresponding vectors 
${\boldsymbol\d}_{1,2,3}$ (see Fig. \ref{fig1}), so their equilibrium: $\sum_\m \varphi_\m(k,q) = 0$, follows from the evident equilibrium: 
$\sum_\m {\boldsymbol\d}_\m = 0$, of these vectors. 
\begin{figure}[h!]
\centering
  \includegraphics[width=9cm]{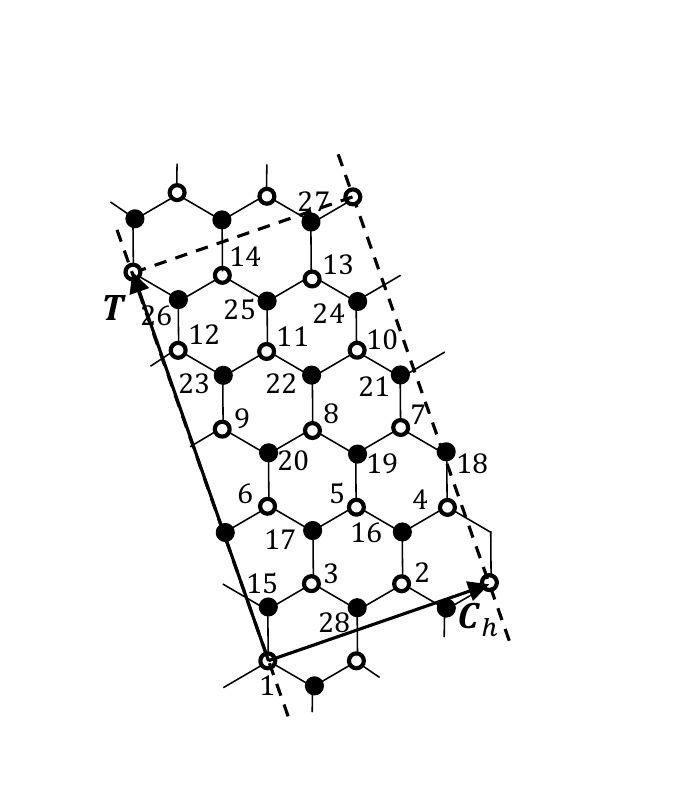}
  \caption{Numeration of atoms within the (2,1) TNT unit cell by Fig. \ref{fig1}, in the order of two sublattices, each in the order of its 
  $t$-coordinates.}
  \label{fig3}
\end{figure}
\begin{figure}[h!]
  \centering
\includegraphics[width=9cm]{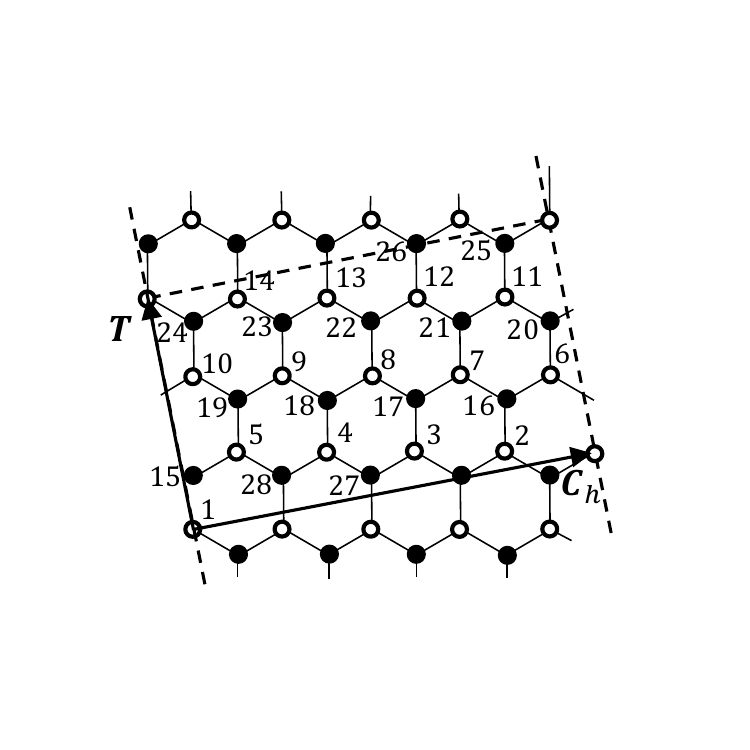}
  \caption{Numeration, similar to that by Fig. \ref{fig3}, for the (4,1) TNT unit cell by Fig. \ref{fig2}.}
\label{fig4}
\end{figure}

The particular positions of each $z_\m(k,q)$ element in this $M$$\times$$M$ matrix depend on the way to enumerate all the $M  \gg 1$ atomic sites 
in the TNT unit cell, creating a huge variety of possible matrix configurations. A certain unification can be reached with the following sublattice-separated 
numeration: 

$\ast$  the first $M/2$ numbers attributed to the atoms of the first sublattice in the increasing order of their $t$ coordinates,

$\ast$ the next $M/2$ numbers follow in the same order for their nearest neighbors from the second sublattice (e.g., at ${\boldsymbol\d}_1$ 
separation, as in the examples of Figs. \ref{fig3}, \ref{fig4}). 

 This brings the Hamiltonian matrix to a more suitable blockwise form:
\be
\hat H(k,q) = \left(\begin{array}{cc}
    0 & \hat h(k,q) \\
    \hat h^\dagger(k,q) & 0
\end{array}\right),
\lb{Hk}
\ee
where the $\hat h(k,q)$ block of $M/2$ dimension and its Hermitian conjugated $\hat h^\dagger(k,q)$ describe the hoppings between two 
sublattices.

Thus, the site numbering as for the $(2,1)$ TNT in Fig. \ref{fig3} leads to a 14$\times$14 submatrix $\hat h(k,q)$ of the multidiagonal form 
explicitly presented by Eq. \ref{h21} (Appendix A), while that for $(4,1)$ TNT in Fig. \ref{fig4} gives it in another multidiagonal form as in 
Eq. \ref{h41}, with their $z_\m(k,q)$ elements defined from Eq. \ref{uvw} for each $n,m$ choice.

Generally, the TNT spectrum consists of $M\n$ 1D modes $\e_{j,k}$, $1 \leq j \leq M\n$, given by the roots of secular equations:
\be
D(\e,k,q) = \det\left(\e\hat 1 - \g\hat H(k,q)\right) = 0.
\lb{det}
\ee
For each $q$ value, this determinant is presented as an $M/2$-degree polynomial of $\e^2$:
\be
D(\e,k,q) = \sum_{j = 0}^{M/2} (-1)^j D_j(k,q)\g^{M - 2j}\e^{2j},
\lb{dek}
\ee
and its roots define a set of $M/2$ positive eigen-energies and of their $M/2$ negative counterparts (from the total of $M\n$ modes). 

It should be noted that, unlike the simpler cases of 2D graphene (2 atomic sites in the unit cell) and ZNTs or ANTs (4 atomic sites in the
unit cell), secular equations for TNTs of $M/2 \geq 14$ orders do not admit general algebraic solutions, thus restricting their treatment 
to numerical calculations or to certain simplifications near special points of their spectra.

 Evidently, all the TNTs beyond the condition by Eq. \ref{met} are insulating, while metallic TNTs (within that condition) contain WMs in their 
 spectra only for the single (from the total of $\n$) transversal wave number $q = 0$ in the Fourier series by Eq. \ref{apj}. For this case, the 
 phases by Eq. \ref{uvw} get simplified to:
\bea
&& \varphi_1(k,0) = \frac{2k}{M}(m + 2l),\nn\\
&& \qquad \varphi_2(k,0) = -\frac{2k}{M}(m + l),\nn\\
&& \qquad\qquad\qquad \varphi_3(k,0) = -\frac{2k}{M}l.
\lb{ml}
\eea

It is readily seen that presence of WM in any CNT spectrum implies the equation for zeroth-order coefficient from Eq. \ref{dek}: 
\be 
D_0(k,0) = 0,
\lb{D0}
\ee
and each its root in $k$ defines a Dirac point crossed by two WMs. So Eq. \ref{D0} permits quantitative detection of Dirac points in the TNT 
spectrum. 

Alternatively, these points follow from a simpler equation:
\be
\sum_{\m = 1}^3 z_\m(k,0) = 0.
\lb{zk0}
\ee
since a zero eigenvalue of $\hat H(k,0)$ implies zero eigenvalue of the submatrix $\hat h(k,0)$, obtained from the $M/2$-vector equation:
\be
\sum_{j' = 1}^{M/2} h_{j,j'}(k,0)\psi_{j'}(k) = 0,
\lb{hpsi}
\ee
for the relative $M/2$-eigenvector $\psi(k)$ of $\hat h(k,0)$. Summation of all the components of this vector equation gives:
\be
\sum_{j = 1}^{M/2}\sum_{j' = 1}^{M/2} h_{j,j'}(k,0)\psi_{j'}(k) = 0.
\lb{sumh}
\ee
Then the multidiagonal structure of $\hat h(k,0)$ brings it to the factored form:
\be
\sum_{\n = 1}^3 z_\n(k,0) \sum_{j = 1}^{M/2}\psi_{j}(k) = 0,
\lb{sum}
\ee
whose solutions should formally come from zeros of each factor. Supposing this factor to be the first sum, as given by Eq. \ref{zk0}, 
one finds Eq. \ref{hpsi} to hold for every its component only with the eigenvector $\psi_j(k) = {\rm const}$ (for all $j$ from a given 
sublattice), which assures the second sum to be nonzero. Contrariwise, if $\sum_{i}\psi_j(k) = 0$ is supposed, then Eq. \ref{hpsi} never 
can hold for every its component, thus suggesting Eq. \ref{zk0} as the simplest check for WM presence.
\begin{figure}[h]
\centering
  \includegraphics[width=9 cm]{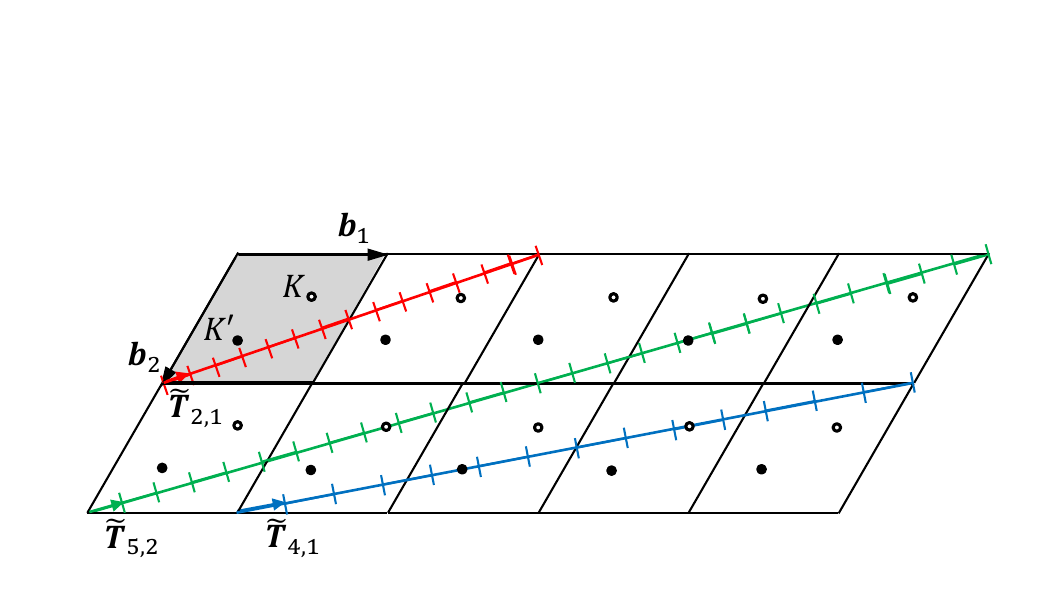}
  \caption{2D graphene BZ (shadowed) with the Dirac points $K$ (white circle) and $K'$ (black circle) together with its multiple replicas. 
  At their overlapping with $M/2 = 14$ multiples of 1D BZ for metallic (4,1) TNT (blue) or with $M/2 = 26$ such multiples for metallic 
  (5,2) TNT (green), each line passes through two Dirac points (white and black circles), but those are never met by the line of $M/2 = 14$ 
  multiples for insulating (2,1) TNT (red).}
  \label{fig5}
\end{figure}

These conditions in some TNTs are illustrated by the plot of their multiple 1D BZs along the relative base vectors $\tilde{\bold T} = 2\pi 
{\bold T}/T^2$, superposed on 2D graphene multiple BZ's (Fig. \ref{fig5}). It shows that, under the condition by Eq. \ref{met}, the 1D BZ
line crosses just one $K$ and one $K'$ points from the 2D BZs at 1/3 and 2/3 of this line length $\pi M/T$ (in $\tilde T$ units).

Thus at the "1/3" crossing, $k = \pi M/(3T)$, we get the phases:
\bea
&& \varphi_1 = \frac{2\pi}{3}(m + 2l), \quad \varphi_2 = -\frac{2\pi}{3}(m + l), \nn\\
&& \qquad\qquad\varphi_3 = -\frac{2\pi}{3}l, 
\lb{var}
\eea
so the resulting $z_\n$ values are defined by the three integers $u_1 = (m + 2l)\, ({\rm mod}\, 3)$, $u_2 = (m + l) \,({\rm mod}\,3)$, 
and $u_3 = l \,({\rm mod}\, 3)$. For any whole $l$ and for any permitted $m\, \neq 0 \,({\rm mod}\, 3)$, the $(u_1,u_2,u_3)$ triple is a 
certain permutation of (0, 1, 2), assuring the condition by Eq. \ref{zk0}. Evidently, the same triples take place for the "2/3" crossing.

\begin{figure}
\centering
  \includegraphics[width=8 cm]{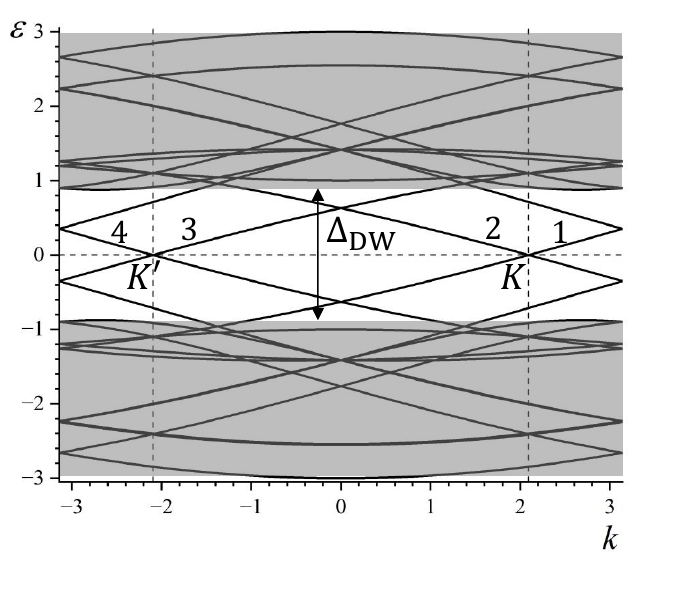}
  \caption{The set of 28 energy bands for $(4,1)$ TNT {\it vs} longitudinal momentum $k$ (in the $1/T$ scale) at transversal number $q = 0$ includes 
  4 WMs (numbered) crossed at Dirac points $K$ and $K'$ (vertical dashed lines). The ranges of resting 24 bands outside the Dirac window $\D_{\rm DW}$ 
  are shadowed.}
  \label{fig6}
\end{figure}

In more detail, the polynomial coefficients in Eq. \ref{dek} calculated for $(4,1)$ TNT are given explicitly in Appendix \ref{Sdet}. Then 
the straightforward calculation of Eq. \ref{D0} using the form of Eq. \ref{41coef} confirms its two roots, $k = \pm 2\pi/3$, as the Dirac points 
(each achieved within a proper multiple of 1D BZ), resulting in 4 WMs within the total of 28 modes, as also seen from the complete numerical 
solution of the secular Eq. \ref{det} in Fig. \ref{fig6}. 
  \begin{figure}
\centering
  \includegraphics[width=10 cm]{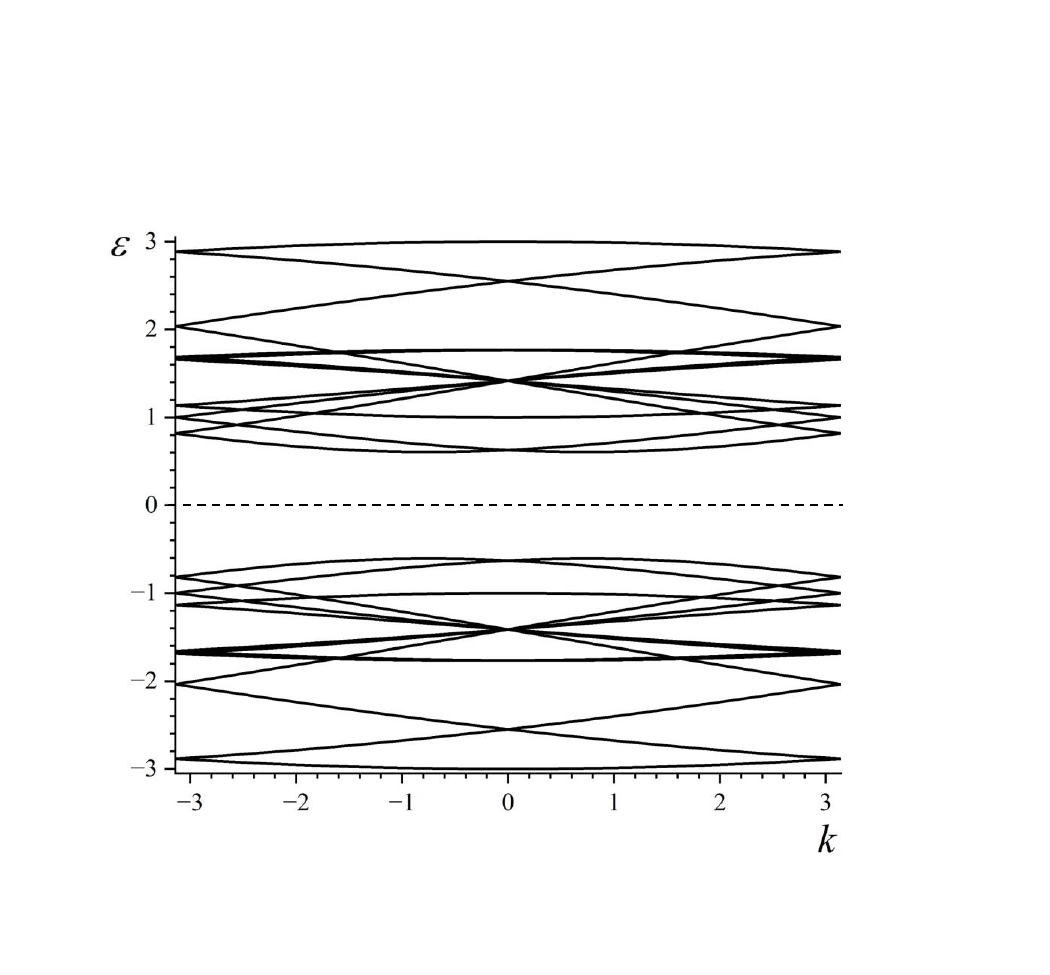}
  \caption{ The set of 28 energy bands for $(2,1)$ TNT {\it vs} $k$ at $q = 0$ does not include WMs.}
  \label{fig7}
\end{figure}

The remaining 24 modes occupy the energy range outside the Dirac window of width $\D_{\rm DW}$ around the Fermi level, also shown in Fig.
\ref{fig6}. Within this window, the full TNT spectrum can be restricted to only its four WMs with their low-energy dispersion approximated 
to purely linear.

This can be compared to the case of $(2,1)$ TNT (with no WMs expected); here the secular determinant at $q = 0$ again takes the form of Eq. \ref{dek} 
but with different coefficients, as given in Appendix B. The same check by Eq. \ref{D0} for (2,1) TNT with the use of Eq. \ref{21coef} readily 
shows that here $D_0(k,0)$ never turns zero along the whole $k$ range, thus excluding WMs in this TNT spectrum, also seen from its complete 
numerical solution in Fig. \ref{fig7}. 

The next treatment of low-energy electronic spectra and impurity effects on them is focused on WMs with their two practical parameters: the 
relative spectral weight $w = 4/(M\n)$ and the Fermi velocity $v_{\rm F}$. Together, they define the important locator function:
\be
G_0(\e) = \frac{1}{M\n N}\sum_{j,k} \frac{1}{\e - \e_{j,k}},
\lb{loc}
\ee
and, in the adopted approximation of linear WMs, this function results a purely imaginary constant: $G_0(\e) = ig_0$, with $g_0 = wT/(2\hbar 
v_{\rm F})$ decisive for the local impurity level broadening. 

The Fermi velocity in a given TNT can be found analytically from the first two terms in the series by Eq. \ref{dek}: 
\be
v_{\rm F} = \frac{\g}{\hbar}\sqrt{\left.\frac{\partial^2D_0(k,0)/\partial k^2}{2D_1(k,0)}\right|_{k = K}}.
\lb{vf}
\ee
Thus, for the case of (4,1) TNT with $T = \sqrt 7 a$ and the functions $D_0(k,0), D_1(k,0)$ given by Eq. \ref{41coef}, the result by Eq. 
\ref{vf} (in $\g a^2/(\hbar T)$ units) reproduces exactly the standard graphene value $\sqrt 3/2$ (in $\g a/\hbar$ units). Then the locator
value for this TNT results simply $g_0 = 1/(\sqrt 3 \g)$. 

When going to TNTs of yet higher $(n,m)$ indices, full analytic forms of their secular determinants, like Eq. \ref{41coef}, turn already 
difficult for available calculation sources. But an easier definition of $v_{\rm  F}$ follows from the direct secular equation: 
\be
\g\sum_{j'} h_{1,j'}(k,0)\psi_{j'}(k) = \e_{1,k}\psi_1(k)
\lb{sec}
\ee
(here for the WM index $j = 1$ as in Fig. \ref{fig6}, but the same will result for all other WMs). Then, expanding the non-zero elements 
$z_\n(k,0)$ of the $\hat h(k,0)$ matrix near the Dirac point $K$ up to linear terms in $\k = k - K$, approximating the eigenvector components 
by their constant value: $\psi_1(k) \approx \psi_1(K) = {\rm const}$, and taking the WM eigen-energy as $\e_{1,k} \approx \hbar v_{\rm F}(k - K)$, 
we get the Fermi velocity as:
\be
v_{\rm F} = \frac \g\hbar\left|\sum_{\n = 1}^3 \frac{dz_\n(k,0)}{dk}\right|_{k = K},
\lb{vfd}
\ee
which can be readily calculated with use of Eq. \ref{ml} for any TNT satisfying the Eq. \ref{met} condition. The crucial result of this 
calculation for all WMs is:
\be
v_{\rm F} = \frac{\sqrt 3\g a^2}{2\hbar T} 
\ee
 (see Appendix C), generalizing the above particular result from Eq. \ref{vf}. This permits to express the locator constant for any metallic CNT as: 
\be
g_0 = \frac{1}{\sqrt 3  \g \n},
\lb{g0}
\ee
that is simply proportional to the local density $1/\n$ of the standing ($q = 0$) transversal wave. 

Now we are in a position to consider perturbations of low-energy electronic spectra in such TNTs under disorder from impurity centers. 

\section{Impurity effects on WMs in twisted nanotubes}\lb{Imp}

Various disorder effects on carbon nanosystems are being studied both experimentally \cite{Jalili, Araujo, Vejpravova} and theoretically \cite{Wakabayashi1996, Nakada1996, Wehling2009,Skrypnyk,Irmer2018,Pogorelov2020,PKL2021, PL2022}. In particular, important effects on 
their electronic spectra are expected from $d$-metal (Cu, Ag, Au) impurity adatoms \cite{Castro Neto, Amft, Faye} and for the case of 
TNT's they can be treated in a similar way to the previous analysis in simpler ANTs and ZNTs \cite{PL2024}. 
\begin{figure}[h!]
\centering
  \includegraphics[width=8 cm]{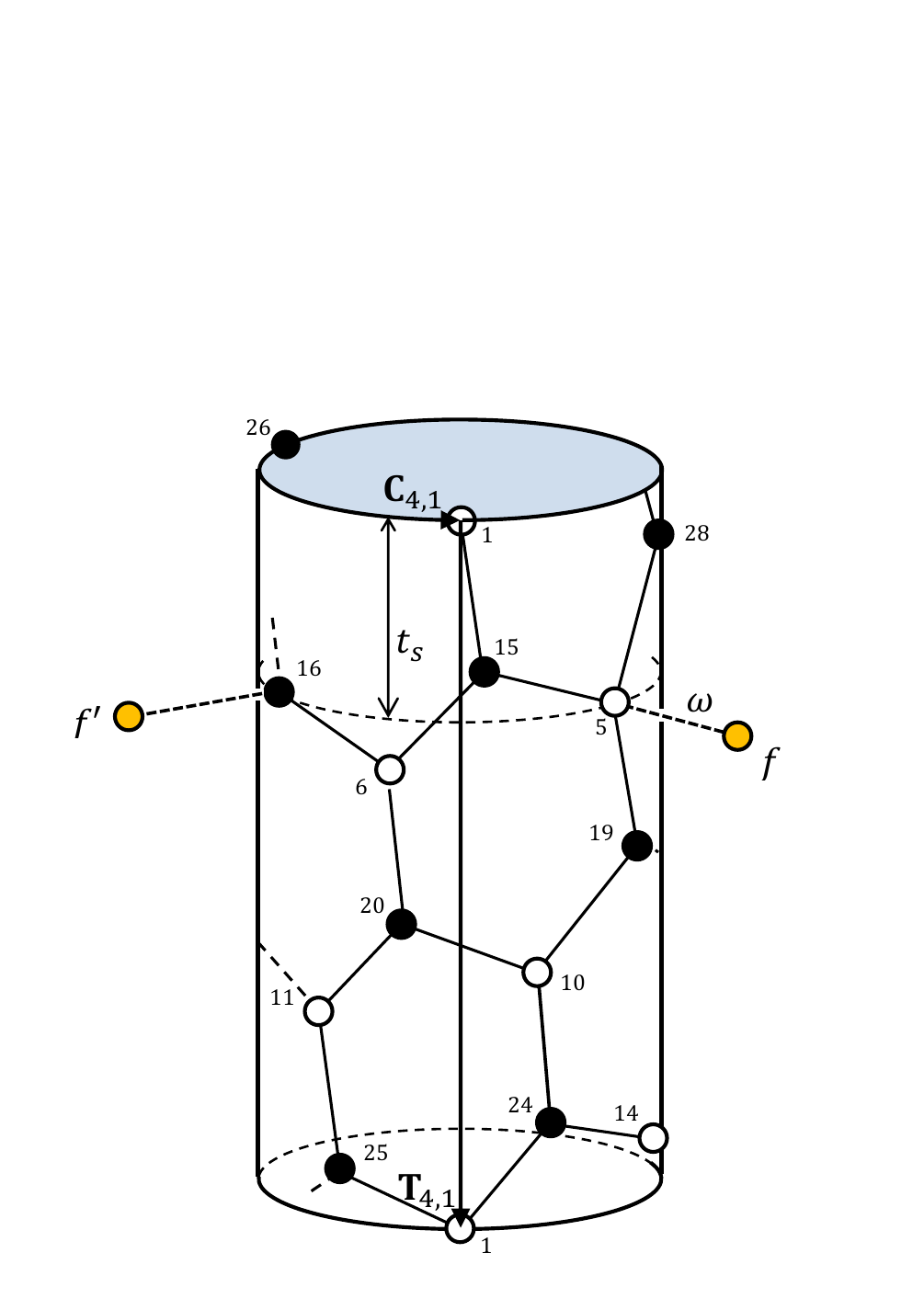}
  \caption{The cylindrical unit cell of the (4,1) TNT with atoms numeration as in Fig. \ref{fig4}. The impurity adatoms (yellow circles) 
  are linked through the hybridization constant $\o$ to the nearest carbon atoms. Note the coincidence of longitudinal coordinates, $t_f = 
  t_{f'}$, for the sites $f = 5$ and $f' = 16$ (from different sublattices).}
  \label{fig8}
\end{figure}

First of all, we reduce the TNT spectrum to its four WMs with the mode indices as in Fig. \ref{fig3} and combine them into the 4-spinor:
\be
\Psi_k = \left(\begin{array}{c}
     \a_{k,1}  \\
    \a_{k,2}  \\
    \a_{k,3}  \\
    \a_{k,4}  
\end{array}\right)
\ee

Then the WM-reduced TNT Hamiltonian is written as:
\be
H_{WM} = \int_{-\pi}^\pi dk \Psi_k^\dagger \hat h_k\Psi_k ,
\ee
where the 4$\times$4 matrix:
\be
\hat h_k = \left(\begin{array}{cc}
     \e_{k,K} \hat \t_3 & 0 \\
    0 & \e_{k,K'} \hat \t_3 
\end{array}\right),
\lb{hk}
\ee
includes 2$\times$2 Pauli $\hat \t_3$ matrices. Using the result of Eq. \ref{vfd}, its eigenenergies are linearized near the Dirac
points as $\e_{k,K} \approx v_{\rm F}(k - K)$ and $\e_{k,K'} \approx v_{\rm F}(k - K')$. 

For impurity adatoms located over random TNT sites $f$, we apply the $s$-$d$ hybrid Anderson model (AM) \cite{Anderson1961} with its two
perturbation parameters: impurity (low energy) level $\e_{res}$ and the constant $\o$ of its (weak) hybridization with the closest carbon
atoms (Fig. \ref{fig11}), but, as in previous studies \cite{PKL2021, PL2022, PL2024}, omitting the Hubbard term on site $U$. As was
shown there, this two-parametric model is more adequate to the considered situation than simpler one-parametric models, such as the Anderson
model of random on-site energy shift \cite{Anderson1958} or the Lifshitz model of fixed energy shift on random sites \cite{Lifshitz1964}. 

Within the chosen AM framework, the perturbation Hamiltonian is written in terms of local impurity operators $\b_f$ and spinors $\Psi_k$ as: 
\bea 
H_{AM} & = & \sum_f \left[\e_{res} \b_f^\dagger\b_f \right. \nn\\
& + & \left.\frac{\o}{\sqrt{2\pi}}\int_{-\pi}^\pi dk ({\rm e}^{ikt_f}\b_f^\dagger u_f^\dagger\Psi_{k}  + {\rm h.c.})\right], 
\lb{pert}
\eea
where $t_f$ is the $f$ site longitudinal coordinate and the row-spinors $u_f^\dagger = \pm(i,i,i,i)/{\sqrt M}$ relate to the eigen-vectors 
of WM's with $k = K, K'$ at $f$ site (see after Eq. \ref{sum}).

The next treatment is done as usually in terms of two-time (advanced) Green functions (GFs), Fourier transformed in energy 
\cite{Zubarev1960,Bonch, Economou1979}: 
\be
\langle\langle A|B\rangle\rangle_\e = \frac i\pi \int_{-\infty}^0 {\rm e}^{i(\e - i0)t}\langle \left\{A(t),B(0)\right\}\rangle dt,
\ee
including the Heisenberg representation operators $O(t) = {\rm e}^{i H t}O{\rm e}^{-i H t}$ with the Hamiltonian $H$, their grand-canonical 
statistical average:
\[\langle O(t)\rangle = \frac{{\rm Tr}\,\left[{\rm e}^{-\b(H - \m)}O(t)\right]}{{\rm Tr}\,\left[{\rm e}^{-\b(H - \m)}\right]},\] 
with inverse temperature $\b$ and chemical potential $\mu$, and the anti-commutator $\{.,.\}$ of fermion operators. These GFs are calculated
from the equation of motion:
\be
\e \llang A|B\rrang_\e = \langle\left\{A(0),B(0)\right\}\rangle + \llang [A,H]|B\rrang_\e,
\lb{de}
\ee
with the commutator $[.,.]$. In what follows the energy subindex at GFs is mostly omitted (or enters directly as its argument).

Consider the GF matrix $\hat G_{k,k'} = \langle\langle\Psi_{k}|\Psi_{k'}^\dagger\rangle\rangle$ describing correlations between the WM states. 
Here Eq. \ref{de} with the Hamiltonian $H = H_{WM} + H_{AM}$ takes the form:
\bea
\hat G_{k,k'} & = & \hat G_k^{(0)}\d(k - k')\nn\\
& + & \frac{\o}{\sqrt{2\pi}}\sum_f {\rm e}^{-ikt_f} \hat G_k^{(0)} \langle\langle\b_f|\Psi_{k'}^\dagger\rangle\rangle u_f,
\lb{gf}
\eea
with the unperturbed GF matrix $\hat G_k^{(0)} = (\e - \hat h_k)^{-1}$, the Dirac $\d$-function, and involving the mixed GFs $\langle\langle\b_f|\Psi_{k'}^\dagger\rangle\rangle$  (between impurity and WM states). 

Now continue the chain of equations of motion, separating at each next step the terms with GFs already present at previous steps. Thus, for the 
mixed GF in the r.h.s. of Eq. \ref{gf} we obtain:
\bea
&&\langle\langle\b_f|\Psi_{k'}^\dagger\rangle\rangle (\e - \e_{res}) = \frac{\o}{\sqrt{2\pi}} u_f^\dagger\left\{{\rm e}^{ik't_f} \hat G_{k',k'}\right.\nn\\ 
&& + \, {\rm e}^{ikt_f} \hat G_{k,k'} + \int_{-\pi}^\pi dk'' \left[1 - \d(k'' - k) - \d(k'' - k')\right]\nn\\
&& \qquad\qquad\qquad\qquad\qquad\qquad\left.\times\, {\rm e}^{ik''t_f}\hat G_{k'',k'}\right\},
\lb{fk}
\eea
with separated the most important momentum-diagonal $\hat G_{k',k'}$ and the initial $\hat G_{k,k'}$ terms. For the resting terms, the chain is
continued as:
\bea
&& \hat G_{k'',k'} = \frac{\o}{\sqrt{2\pi}}\hat G_{k''}^{(0)}({\rm e}^{-ik''t_{f}}  \langle\langle\b_{f}|\Psi_{k'}^\dagger\rangle\rangle u_f\nn\\
&& + \sum_{f' \neq f} {\rm e}^{-ik''t_{f'}} \langle\langle\b_{f'}|\Psi_{k'}^\dagger\rangle\rangle u_f').
\lb{k'k}
\eea
Here the separated $\langle\langle\b_{f}|\Psi_{k'}^\dagger\rangle\rangle$ term generates the impurity self-energy $\Sigma_0 = \o^2 G_0$ for 
the mixed GF within Eq. \ref{fk}. It is then rewritten as:
\bea
&&\langle\langle\b_f|\Psi_{k'}^\dagger\rangle\rangle (\e - {\tilde \e}_{res}) = \frac{\o}{\sqrt {2\pi}}u_f^\dagger\left\{{\rm e}^{ik't_f} \hat 
G_{k',k'}\right.\nn\\
&& + \,{\rm e}^{ikt_f} \hat G_{k,k'} +\, \int_{-\pi}^\pi dk'' \left[1 - \d(k'' - k) - \d(k'' - k')\right]\nn\\
&& \qquad\qquad\qquad\qquad\qquad\qquad \left. \times\, {\rm e}^{ik''t_f} \hat G_{k'',k'}\right\},
\lb{fk1}
\eea
with the renormalized resonance energy ${\tilde \e}_{res} = \e_{res} + \S_0$. 

Now the mixed GFs can be eliminated from Eq. \ref{gf} to present it as WM scattering by the effective impurity potential:
\bea
&& \hat G_{k,k'} = \hat G_k^{(0)}\left\{\d(k - k') + \frac{c\o^2}{\e - {\tilde \e}_{res}}\hat G_{k',k'}\right. \nn\\
&& + \frac{1}{\sqrt{2\pi}}\sum_f\int_{-\pi}^\pi dk''\left[1 - \d(k'' - k) - \d(k'' - k')\right]\nn\\
&& \qquad\qquad\left. \times\,{\rm e}^{i(k'' - k)t_f}\frac{\o^2}{\e - {\tilde \e}_{res}} u_f^\dagger\hat 
G_{k'',k'}u_f\right\},
\lb{gfd}   
\eea
with the impurity concentration $c = \sum_f (NM)^{-1} \ll 1$.

All the subsequent iterations, restricted to only scatterings at the same impurity site $f$, result in the T-matrix approximated solution for 
the momentum-diagonal GF \cite{ILP1987}:
\be
\hat G_{k,k} \approx \left[\left(\hat G_k^{(0)}\right)^{-1} - c\hat T\right]^{-1},
\lb{tm}
\ee
with the T-matrix (scalar in this case):
\be
\hat T(\e) = \frac{\o^2}{\e - {\tilde \e}_{res}}\hat 1.
\lb{tm1}
\ee
Within the adopted perturbation model, the restructured dispersion law: 
\be
\tilde \e_{k,K} = \e_{k,K} - c\,{\rm Re}\,T(\tilde\e_{k,K}),
\lb{rd}
\ee
 involves the TNT specifics only through the locator value $g_0$ in the renormalized impurity resonance ${\tilde \e}_{res}$.

Now, in analogy with the previous study on impurity induced WM spectrum restructuring in simple ANTs and ZNTs \cite{PL2024}, we can conclude 
about this in all metallic TNTs. 

Generally, such restructuring of an energy band $\tilde \e_{k,K}$ with growing impurity concentration $c$ is driven by the competition between 
WM wave period $2\pi/(v_{\rm F}|k - K|)$ and its lifetime $\t_{1,k}$, that is between propagation and localization. This is controlled by the 
Ioffe-Regel criterion for a quasiparticle state to remain conductive \cite{IoffeRegel}:
\be
k\frac{\partial \tilde\e_{k,K}}{\partial k}  \gtrsim \t_{j,k}^{-1},
\lb{IR}
\ee
where the inverse lifetime in the considered case: $\t_{1,k}^{-1} = \o^2g_0$. The threshold form of Eq. \ref{IR} when $\gtrsim$ turns $=$ is 
used to estimate the onset of Mott's mobility edges $\e_{mob}$ \cite{Mott} between conducting and localized ranges in the spectrum. 
\begin{figure}
\centering
  \includegraphics[width=6 cm]{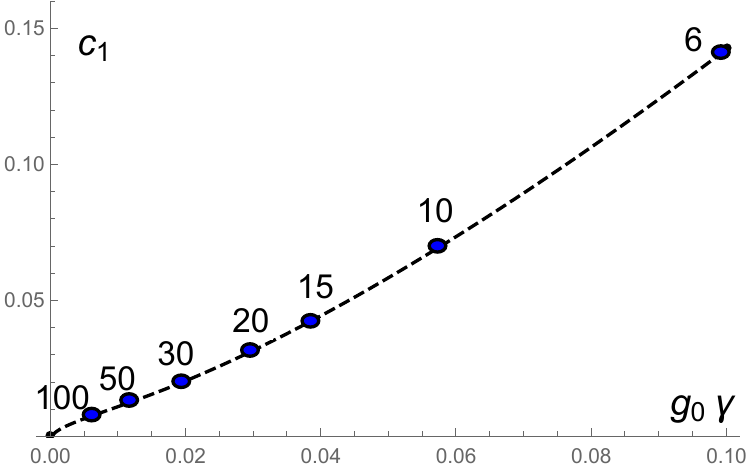}
  \caption{Critical concentrations $c_1$ of Cu impurities in different TNTs (blue circles labeled by $\n$-numbers) in function of the related
  $g_0$ parameter, fitted by Eq. \ref{c1} (dashed line).}
  \label{fig9}
\end{figure}
For doped CNTs, the indicated competition develops with growing $c$ to produce qualitative restructurings of initial WM 
spectrum, marked by two critical values of $c$ \cite{PL2022}. 

The lower one, $c_0$, indicates emergence of a range of localized states (mobility gap) around the impurity resonance level $\e_{res}$ at $c > c_0$, known as incoherent restructuring of spectrum \cite{ILP1987}. This critical value:
\be
c_0 = \frac{(\o g_0)^2}{2}\left(1 + \sqrt{1 + 4\frac{\e_{res}}{\o^2g_0}}\right),
\lb{c0}
\ee
follows from the explicit solution of the threshold Eq. \ref{IR} at $\e = \e_{res}$.

\begin{figure}
\centering
  \includegraphics[width= 4cm, angle=90]{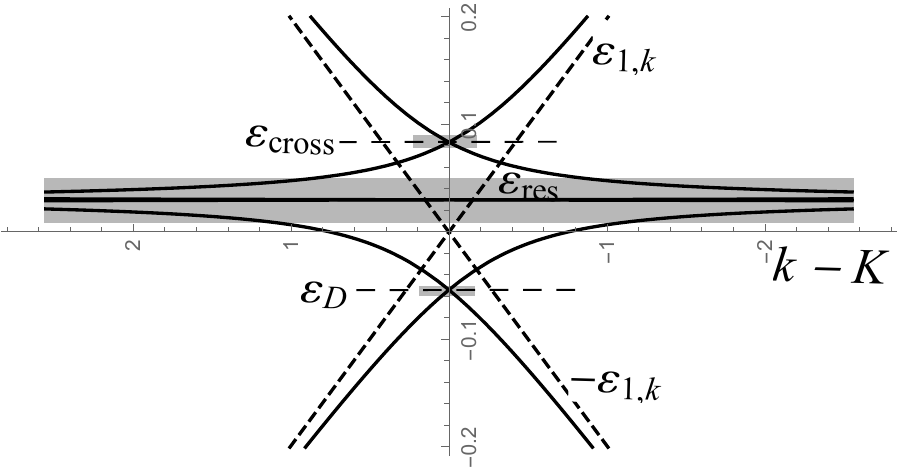}
   \caption{Intermittence of conducting and localized (shadowed) ranges in the WM spectrum of TNT with $\n = 20$ under 
   effect of Cu impurities with concentration $c = 0.05 > c_1 \approx 0.032$. The broader mobility gap around $\e_{res}$ is surrounded by 
   conducting ranges (with their velocities $v \ll v_{\rm F}$), delimited by narrow mobility gaps around $\e_{cross}$ and $\e_{\rm D}$.}
\label{fig10}
\end{figure} 

With further $c$ growth, it can reach another critical value when the resonance splitting between WM and $\e_{res}$ surpasses the initial mobility 
gap to open an additional conducting range within it,  known as coherent restructuring of spectrum \cite{ILP1987}. This $c_1$ value follows from 
the numerical solution of Eq. \ref{IR} and its $g_0$-dependence for the instance of $\e_{res} = 0.03\g$ and $\o = 0.3\g$ (as for Cu impurities 
\cite{PKL2021}) is well fitted by: 
\be
c_1 \approx \sqrt[3]{0.01 g_0^2 + 28 g_0^4},
\lb{c1}
\ee
illustrated in Fig. \ref{fig9} for a $\n$-series of TNTs. 

Such a splitting creates an additional mobility gap around the WMs crossing point $\e_{cross}$ (Fig. \ref{fig10}), due to the vanishing 
WM momentum $k - K \to 0$ at $\e \to \e_{cross}$, and an additional conductive range below this gap, with reduced velocity of its states $v 
\ll v_{\rm F}$ (like "heavy fermions") and with their amplitudes decayed away from impurity sites. 

\begin{figure}
\centering
  \includegraphics[width=7cm]{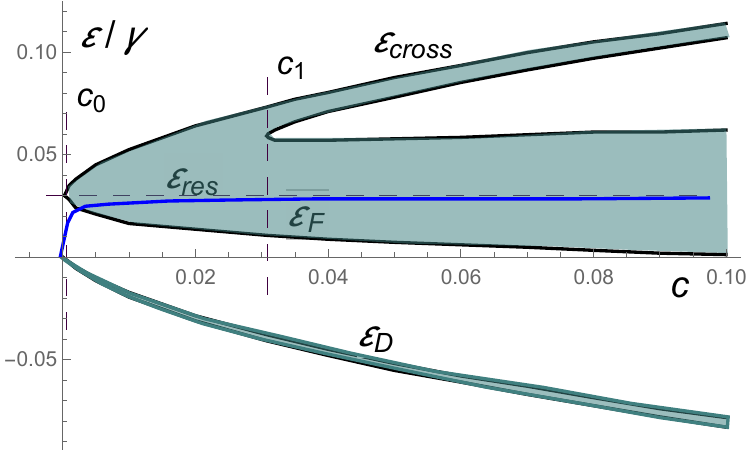}
   \caption{Restructuring of the WM spectrum of TNT as in Fig. \ref{fig10} with growing Cu impurity concentration $c$: rise of the Fermi level 
   $\e_{\rm F}$ (blue line) until staying fixed near the impurity resonance level $\e_{res}$, opening of a mobility gap (shadowed) around $\e_{res}$ 
   at $c > c_0 \approx 3\cdot 10^{-4}$ and its widening until being split at $c > c_1 \approx 3.1\cdot 10^{-2}$ into a narrow conducting range and 
   a narrow mobility gap around $\e_{cross}$. A narrow mobility gap around the shifted Dirac level $\e_{\rm D}$ exists at all $c > 0$.}
\label{fig11}
\end{figure}

The intermittence of narrow conducting and localized ranges in the coherently restructured spectrum could be promising for highly sensitive 
external controls on doped CNT conductance, this sensitivity growing with $\n$ values. However, this growth gets restricted by the opposite 
effect of narrowing the Dirac window for WMs with $\n$ as $\D_{\rm DW} \sim 1/\n$ \cite{PL2024}. Then, in order to retain $\D_{\rm DW}$ 
over the coherent restructuring range around $\e_{res}$ (not to perturb it), the $\n$ growth should be delimited. Thus, for the above instance 
of Cu impurities, this limit is estimated as $\n \lesssim 60$. Nevertheless, as seen from Fig. \ref{fig9}, this limit still permits sensitive 
controls of conductance in properly chosen CNTs at suitably low doping levels of few percents. 

\section{Beyond T-matrix}\lb{Bey}

We can yet briefly discuss the extension of the T-matrix approximation by Eq. \ref{tm} to the more general self-energy form: 
\be
(\hat G_k)^{-1} = (\hat G_k^0)^{-1} - \hat \S_k,
\lb{Gk}
\ee
with the self-energy matrix $\hat\S_k$ presented by the group expansion (GE) \cite{ILP1987}:
\be
\hat\S_k = c\hat T\bigl(\hat 1 + c \hat B_k + \dots\bigr).
\lb{GE}
\ee
Here the sum:
\be
\hat B_k = \sum_{r \neq 0} \bigl[{\rm e}^{-ikt_r}\hat A_{t_r} + \hat A_{t_r}^2\bigr]\bigl[\hat 1 - \hat A_{t_r)}^2\bigr]^{-1}
\lb{Bk}
\ee
describes the effect of multiple scatterings by a pair of impurities at difference $r$ of their positions and longitudinal distance 
$t_r$ between them through the related scattering matrix $\hat A_t = \hat T G_t$. Notably, the correlator function:
\be
G_t = \frac{1}{4M}\int_{-\pi}^\pi dk'\left[1 - \d(k - k')\right] {\rm e}^{ik't}u^\dagger \hat G_{k'}^{(0)}u,
\lb{Gt}
\ee
extends the above discussed locator function $G_0$ to a finite distance $t$ (with the important possibility of zero longitudinal 
distance $t_r = 0$ in the sum by Eq. \ref{Bk}, see Fig. \ref{fig8}). The omitted terms in 
the r.h.s. of Eq. \ref{GE} correspond to contributions by clusters of three and more impurities.
	
However, by analogy with the previous study on disordered ZNTs and ANTs \cite{PL2024}, the GE corrections to the above T-matrix 
result are negligible for disordered TNTs (beyond a very narrow vicinity of $\e_{res}$ deeply within the localized range) and thus 
do not change the general composition of their spectra. Also these simple T-matrix results are found in a good agreement with their 
another extension to the self-consistent GF form \cite{Freed}. 

\section{Discussion and conclusions}\lb{Disc}
The given analysis demonstrates either similarities and differences of impurity disorder effects on WMs in various types of carbon 
nanotubes, to contribute into the general context of conductance in 1D-like disordered systems \cite{Ando1998,Ando2000,Vosk}. It indicates 
importance of each nanotube topology (besides only its size) for the detailed electronic structure, especially at low energies. The known 
earlier qualitative difference between metallic CNTs (with strong restructuring of their WMs under low impurity doping) and insulating ones 
(practically insensitive to such doping) can be now yet  completed by the comparison between different metallic CNTs in their sensitivity to 
impurity dopants, based on the values of critical concentration $c_1$ by  Eq. \ref{c1} and locator $g_0$ by Eq. \ref{g0}. 

This comparison readily shows that, for the same CNT diameter $D$, the highest sensitivity (the lowest $g_0$) at the widest $\D_{\rm DW}$ 
is reached in ZNT:
\be 
g_0 = \frac{a}{\sqrt 3 \pi D\g},\qquad\qquad \D_{\rm DW} \approx \frac{2\sqrt 3 a}{D}\g,
\lb{Z}
\ee
 followed by ANT:
\be 
g_0 = \frac{a}{\pi D\g},\qquad\qquad \D_{\rm DW} \approx \frac{2 a}{D}\g,
\lb{A}
\ee
and only then by all TNTs: 
\bea
&& g_0 = \frac{T}{\pi D\g},\qquad \D_{\rm DW} \approx \frac {1.75a}{D}\g. 
\lb{T}
\eea
 This suggests the choice for doped ZNTs as the most prospective triggering elements in future nanotechnologies. 

Finally, yet the validity of the adopted approximation for CNT Hamiltonian by Eq. \ref{HZ} can be discussed. As seen from the known analysis on metallic 
and insulating ZNTs and ANTs \cite{Barnett}, the first principles corrections to the simple zone-folding results can modify the initially 
gapped spectra (but far from the Dirac points), however they have practically no effect on zero-gap WMs, like those of our central interest 
here, so justifying (at least, qualitatively) the above conclusions about impurity effects on their spectra. 

Under these effects, the most important factors of CNT effectiveness are the extremely fine intervals between conducting and localized energy 
ranges (as sen in Fig. \ref{fig10}) and also the Fermi level fixing  near $\e_{res}$ at all $c > c_0$ \cite{PL2022,PL2024} (as seen in Fig. \ref{fig11}). Both them permit, at $c > c_1$,  repeated insulator/metal and metal/insulator transitions at {\it monotonous} variation 
of external control (within a few mV scale). Also, there can be various types of these controls (electric and magnetic fields, e.m. radiation, 
etc.) used in such high sensitivity devices with properly chosen CNTs.

Possible practical applications of doped CNTs electronic properties can be sought first of all in the field of controlled conductance 
devices, analogs to the traditional semiconducting diodes and their derivatives but with qualitatively higher levels of sensitivity and 
compactness. Otherwise, they can also provide new perspectives for high density memory elements \cite{Ann2018, Fan2024}, chemical/organic 
sensing elements \cite{Pumera}, (infrared) radiation receivers and transformers, supercapacitors, and many others.

\section{Acknowledgements}

The authors are grateful to V.P. Gusynin, A.A. Eremko and S.G. Sharapov for a substantial discussion of different aspects of the present 
work. The work of VT was supported in part by United States Department of Energy under Grant No DE-FG02-07ER46354, VML acknowledges the 
support from Grant ID 1290587 of the Simons Foundation (USA).

\appendix

\section{TNT matrices}\lb{Mat}

Using the position numbering as for the $(4,1)$ TNT in Fig. \ref{fig4} leads to a 14$\times$14 submatrix $\hat h$ of the 
multi-diagonal form (with the $k,q$ arguments omitted for brevity):
\bea 
&&\hat h =\nn\\
&& \left(\begin{array}{cccccccccccccc}
        z_1 & 0 & 0 & 0 & 0 & 0 & 0 & 0 & 0 & z_2 & z_3 & 0 & 0 & 0\\
        0 & z_1 & 0 & 0 & 0 & 0 & 0 & 0 & 0 & 0 & z_2 & z_3 & 0 & 0 \\ 
        0 & 0 & z_1 & 0 & 0 & 0 & 0 & 0 & 0 & 0 & 0 & z_2 & z_3 & 0 \\
        0 & 0 & 0 & z_1 & 0 & 0 & 0 & 0 & 0 & 0 & 0 & 0 & z_2 & z_3 \\
         z_3 & 0 & 0 & 0 & z_1 & 0 & 0 & 0 & 0 & 0 & 0 & 0 & 0 & z_2 \\
         z_2 & z_3 & 0 & 0 & 0 & z_1 & 0 & 0 & 0 & 0 & 0 & 0 & 0 & 0 \\
         0 & z_2 & z_3 & 0 & 0 & 0 & z_1 & 0 & 0 & 0 & 0 & 0 & 0 & 0 \\
         0 & 0 & z_2 & z_3 & 0 & 0 & 0 & z_1 & 0 & 0 & 0 & 0 & 0 \\
         0 & 0 & 0 & z_2 & z_3 & 0 & 0 & 0 & z_1 & 0 & 0 & 0 & 0 & 0 \\
         0 & 0 & 0 & 0 & z_2 & z_3 & 0 & 0 & 0 & z_1 & 0 & 0 & 0 & 0 \\
         0 & 0 & 0 & 0 & 0 & z_2 & z_3 & 0 & 0 & 0 & z_1 & 0 & 0 & 0 \\
         0 & 0 & 0 & 0 & 0 & 0 & z_2 & z_3 & 0 & 0 & 0 & z_1 & 0 & 0 \\
         0 & 0 & 0 & 0 & 0 & 0 & 0 & z_2 & z_3 & 0 & 0 & 0 & z_1 & 0 \\
         0 & 0 & 0 & 0 & 0 & 0 & 0 & 0 & z_2 & z_3 & 0 & 0 & 0 & z_1
    \end{array}\right),\nn\\
    &&
    \lb{h41}
    \eea
with $z_1 = {\rm e}^{i3k/14}$, $z_2 = {\rm e}^{-ik/7}$, $z_3 = {\rm e}^{-ik/14}$. 

Otherwise, using the numbering for $(2,1)$ TNT as in Fig. \ref{fig3} gives such submatrix as:
\bea 
&&\hat h =\nn\\
&& \left(\begin{array}{cccccccccccccc}
        z_1 & 0 & 0 & 0 & 0 & 0 & 0 & 0 & 0 & 0 & 0 & z_2 & z_3 & 0\\
        0 & z_1 & 0 & 0 & 0 & 0 & 0 & 0 & 0 & 0 & 0 & 0 & z_2 & z_3 \\ 
        z_3 & 0 & z_1 & 0 & 0 & 0 & 0 & 0 & 0 & 0 & 0 & 0 & 0 & z_2 \\
        z_2 & z_3 & 0 & z_1 & 0 & 0 & 0 & 0 & 0 & 0 & 0 & 0 & 0 & 0 \\
        0 & z_2 & z_3 & 0 & z_1 & 0 & 0 & 0 & 0 & 0 & 0 & 0 & 0 & 0 \\
        0 & 0 & z_2 & z_3 & 0 & z_1 & 0 & 0 & 0 & 0 & 0 & 0 & 0 & 0 \\
         0 & 0 & 0 & z_2 & z_3 & 0 & z_1 & 0 & 0 & 0 & 0 & 0 & 0 & 0 \\
         0 & 0 & 0 & 0 & z_2 & z_3 & 0 & z_1 & 0 & 0 & 0 & 0 & 0 \\
         0 & 0 & 0 & 0 & 0 & z_2 & z_3 & 0 & z_1 & 0 & 0 & 0 & 0 & 0 \\
         0 & 0 & 0 & 0 & 0 & 0 & z_2 & z_3 & 0 & z_1 & 0 & 0 & 0 & 0 \\
         0 & 0 & 0 & 0 & 0 & 0 & 0 & z_2 & z_3 & 0 & z_1 & 0 & 0 & 0 \\
         0 & 0 & 0 & 0 & 0 & 0 & 0 & 0 & z_2 & z_3 & 0 & z_1 & 0 & 0 \\
         0 & 0 & 0 & 0 & 0 & 0 & 0 & 0 & 0 & z_2 & z_3 & 0 & z_1 & 0 \\
         0 & 0 & 0 & 0 & 0 & 0 & 0 & 0 & 0 & 0 & z_2 & z_3 & 0 & z_1
    \end{array}\right).\nn\\
    &&
    \lb{h21}
    \eea
with $z_1 = {\rm e}^{i5k/14}$, $z_2 = {\rm e}^{-i2k/7}$, $z_3 = {\rm e}^{-ik/14}$.

\section{Secular determinants}\lb{Sdet}

The straightforward calculation of secular determinant for (4,1) TNT with use of the matrices $\hat h$ by Eq. \ref{h41} (and its 
conjugated $\hat h^\dagger$) gives the coefficients in Eq. \ref{dek} as:
\bea
&& D_0(k,0) = 10892 + 16238 \cos k + 8268 \cos 2k\nn\\
&& \qquad + 1396 \cos 3k + 72 \cos 4k - 2\cos 5k,\nn\\
&& D_1(k,0) = 159040 + 163912\cos k + 45136 \cos 2k\nn\\
&& \qquad\qquad + 4536 \cos 3k + 112 \cos 4k,\nn\\
&& D_2(k,0) = 838614 + 648080 \cos k + 106848 \cos 2k\nn\\
&& \qquad\qquad + 5880 \cos 3k + 42 \cos 4k,\nn\\
&& D_3(k,0) = 2361576 + 1371468 \cos k + 140952 \cos 2k\nn\\
&& \qquad\qquad\qquad + 3892 \cos 3k, \nn\\
&& D_4 (k,0) = 4130182 + 1767192 \cos k + 111650 \cos 2k\nn\\
&& \qquad\qquad\qquad + 1344 \cos 3k,\nn\\
&& D_5(k,0) = 4833192 + 1467452 \cos k + 53592 \cos 2k\nn\\
&& \qquad\qquad\qquad + 28 \cos 3k,\nn\\
&& D_6(k,0) = 3938452 + 804776 \cos k + 14980 \cos 2k, \nn\\
&& D_7(k,0) = 2281180 + 20852 \cos k + 2188 \cos 2k,\nn\\
&& D_8(k,0) = 946358 + 68684 \cos k + 126 \cos 2k, \nn\\
&& D_9(k,0) = 280420 + 9996 \cos 2k,\nn\\
&& D_{10}(k,0) = 58548 + 812 \cos k, \nn\\
&& D_{11}(k,0) = 8372 + 28 \cos k, \nn\\ 
&& D_{12}(k,0) = 777, \quad D_{13}(k,0) = 42, \nn\\
&& D_{14}(k,0) = 1.
\lb{41coef}
\eea

A similar procedure for (2,1) TNT with use of the matrices by Eq. \ref{h21} (having the same dimension) gives another set of 
coefficients for Eq. \ref{dek}:
\bea
&& D_0(k,0) = 44884 - 8478 \cos k + 460 \cos 2k -  2\cos 3k,\nn\\
&& D_1(k,0) = 419104 - 47488 \cos k + 1120 \cos 2k,\nn\\
&& D_2(k,0) = 1702568 - 105056 \cos k + 952 \cos 2k,\nn\\
&& D_3(k,0) = 3998064 - 120512 \cos k + 336 \cos 2k,\nn\\
&& D_4 (k,0) = 6087326 - 77000 \cos k + 42 \cos 2k,\nn\\
&& D_5(k,0) = 6380416 - 25984 \cos k,\nn\\
&& D_6(k,0) = 4761064 - 2856 \cos k,\nn\\
&& D_7(k,0) = 2574424 - 872 \cos k,\nn\\
&& D_8(k,0) = 1014860 - 308 \cos k, \nn\\
&& D_9(k,0) = 290388 - 28 \cos 2k, \nn\\
&& D_{10}(k,0) = 59360,\qquad D_{11}(k,0) = 8400, \nn\\
&& D_{12}(k,0) = 777,\qquad D_{13}(k,0) = 42,\nn\\
&& D_{14}(k,0) = 1.
\lb{21coef}
\eea
This difference is completely due to different phases $\varphi_j$ of their matrix elements $z_j$ as indicated after Eqs. \ref{h21}, 
\ref{h41}, resulting from the difference in chiral angles $\theta$ for these two TNT's when used in Eq. \ref{uvw}.
\vspace{5pt}

\section{Fermi velocity calculation}\lb{Fv}
Let us consider the explicit secular equation for a WM:
\be
\g\sum_{j' = 1}^{M/2} h_{j,j'}(k) \psi_{j'}(k,0) = \e_k \psi_{j}(k,0).
\lb{sec1}
\ee
By the blockwise structure of $\hat h(k)$ matrix, the $j,j'$ indices in this equation belong to different sublattices, namely: $j$ 
to the 1st sublattice and $j'$ to the 2nd sublattice. Also this sum is restricted to only three non-zero matrix elements $z_\n(k,0)$, 
so its expansion up to linear terms in $\k =k - K$ (with the account of Eq. \ref{zk0}) reads: 
\be
\sum_{\n = 1}^3 z_\n(k,0) = i\sum_{\n = 1}^3 \bigl[z_\n(k,0)\frac{d\varphi_\n(k)}{dk}\bigr]_{k = K}\k + O(\k^2).
\lb{sec2}
\ee
Denoting the constant values $\psi_{j}(K,0)$ for two sublattices as $\psi_1$ and $\psi_2$, we get the relation: $\psi_1 
= i\psi_2$, and, using them as eigenvector components in Eq. \ref{sec1}, obtain:
\be
\e_k = \g\bigl|\sum_{\n = 1}^3 z_\n(k,0)\frac{d\varphi_\n(k)}{dk}\bigr|_{k = K}\k + O(\k^2).
\lb{sec3}
\ee
With the derivatives provided by Eq. \ref{ml}:
\bea
&&\frac{d\varphi_1(k)}{dk} = 2a\frac{m + 2l}{M},\quad \frac{d\varphi_2(k)}{dk} = - 2a\frac{m + l}{M},\nn\\
&&\qquad\qquad\qquad  \frac{d\varphi_3(k)}{dk} = -a\frac{2l}{M},
\lb{d1}
\eea
 the sum in Eq. \ref{sec3} gives:
\be
v_{\rm F} = \frac{2\g a}{\hbar M} \left|m(z_1 - z_2)  + l(2z_1 - z_2 - z_3)\right|.
\lb{vf1}
\ee
Then, using again Eq. \ref{zk0} and also Eq. \ref{nml}, we rewrite this as:  
\be
v_{\rm F} = \frac {2\g a}{\hbar M}\left|n z_1 - m z_2\right|,
\lb{vf2}
\ee
and, with the relations $|z_1| = |z_2| = 1$, ${\rm Re}\,z_1 z_2 = -\frac 12$, finally obtain:
\be
v_{\rm F} = \frac {\sqrt 3}{2}\frac {\g a^2}{\hbar T}, 
\lb{vf3}
\ee
that is just $a/T$ of the standard graphene $(\sqrt 3/2)\g a/\hbar$ value.

\end{document}